\documentclass[lettersize,journal]{IEEEtran}
\usepackage{amsmath,amsfonts}
\usepackage{algorithm}
\usepackage{array}
\usepackage[caption=false,font=normalsize,labelfont=sf,textfont=sf]{subfig}
\usepackage{textcomp}
\usepackage{stfloats}
\usepackage{url}
\usepackage{enumerate}
\usepackage{verbatim}
\usepackage{graphicx}
\usepackage{xcolor}
\usepackage{cite}
\usepackage{multirow}
\usepackage{algorithm}
\usepackage{algpseudocode} 
\usepackage{tabularx,booktabs,textcomp}
\usepackage{amsthm}

\usepackage{amsmath,amssymb,amsthm}

\begin{document}
\title{Regression Based Anomaly Detection in Electric Vehicle State of Charge Fluctuations Through Analysis of EVCI Data}
% \title{Enhancing Cyber Security of Electric Vehicle Charging Infrastructure through Anomaly Detection and Spoofing Classification}

\author{{Sagar Babu Mitikiri$^{a}$, Yash Tiwari$^{b}$, Vedantham Lakshmi Srininvas$^{c}$, Mayukha Pal$^{*d}$}
 %~\IEEEmembership{Member,~IEEE}
 %~\IEEEmembership{Senior Member,~IEEE}
        % <-this % stops a space
\thanks{(Corresponding author: $^{*}$Mayukha Pal)}
\thanks{$^{a}$Mitikiri Sagar Babu is a Data Science Research Intern at ABB Ability Innovation Center, Hyderabad 500084, India and also a Ph.D. Research Scholar at the Department of Electrical Engineering, IIT (ISM) Dhanbad, Dhanbad, 826004, India.}
\thanks{$^{b}$Yash Tiwari is with ABB Ability Innovation Center, Bangalore-560048, IN, working as a Data Scientist}
\thanks{$^{c}$Dr. Vedantham Lakshmi Srinivas is an Asst. Professor in the Department of Electrical Engineering, IIT (ISM) Dhanbad, Dhanbad, 826004, India.}
\thanks{$^{d}$Dr. Mayukha Pal is with ABB Ability Innovation Center, Hyderabad-500084, IN, working as Global R\&D Leader – Cloud \& Analytics (e-mail: mayukha.pal@in.abb.com).}}

%\markboth{Journal of \LaTeX\ Class Files,~Vol.~, No.~, }%

\maketitle

\begin{abstract} With the increase in the number of electric vehicles (EV), there is a need for the development of the EV charging infrastructure (EVCI) to facilitate fast charging, thereby mitigating the EV congestion at charging stations. The role of the public charging station depot is to charge the vehicle, prioritizing the achievement of the desired state of charge (SoC) value for the EV battery or charging till the departure of the EV, whichever occurs first. The integration of cyber and physical components within EVCI defines it as a cyber physical power system (CPPS), increasing its vulnerability to diverse cyber attacks. When an EV interfaces with the EVCI, mutual exchange of data takes place via various communication protocols like the Open Charge Point Protocol (OCPP), and IEC 61850. Unauthorized access to this data by intruders leads to cyber attacks, potentially resulting in consequences like energy theft, and revenue loss. These scenarios may cause the EVCI to incur higher charges than the actual energy consumed or the EV owners to remit payments that do not correspond adequately to the amount of energy they have consumed. This article proposes an EVCI architecture connected to the utility grid and uses the EVCI data to identify the anomalies or outliers present in the EV transmitted data, particularly focusing on SoC irregularities. The proposed methodology involves utilizing a ridge regression based machine learning (ML) model for predicting changes in the SoC. The adversaries have the capability of spoofing these change in SoC values, consequently making the EVCI incapable of achieving the desired task. Three distinct spoofing techniques namely, decimal shifting, incremental array spoofing, and random spoofing are implemented on the data and subsequently tested with the proposed methodology. The results show that the proposed methodology detects the anomaly accurately and also classifies the type of spoofing that causes the anomaly.

\end{abstract}


\section{Introduction}
\label{section: Introduction}
The global electric vehicle (EV) market sales have exhibited robust growth reaching 2 million units sold in the first quarter of 2022, which is a 75\%  increase in sales compared to 2021 \cite{EVsales}. Due to this rising demand, the EV charging infrastructure (EVCI) needs to be developed at the same rate. To reduce the EV charging stations (EVCS) complications in penetration with the grid \cite{ravindran2023novel} and in charging such as charging speed, compatibility, range anxiety, grid power availability, and congestion of EVs at charging stations, etc., These charging stations are located at parking areas of corporate offices, commercial malls, and residential apartments, where there is probability for the vehicle staying in these locations, for the dwell times \cite{pan2020location}. Due to the intermittent power available from the utility grid, charging stations are integrated with energy storage systems like battery energy storage systems (BESS), or renewable energy systems (RES) to meet it's power demand. In addition to their primary function of EV charging, these stations offer auxiliary grid-related services, including the stabilization of frequency and voltage levels, provision of reactive power support, etc., These services require bi-directional power flow that enables the flow of electricity from the grid-to-vehicle (G2V mode) and also from the vehicle-to-grid (V2G mode), particularly during the instances when the demand for the loads in the grid exceeds the generation \cite{erb2010bi}. Furthermore, the charging infrastructure has exhibited consistent augmentation in charging capacity, such as ultra-fast direct current (DC) \cite{aggeler2010ultra} and extreme fast charging (XFC) \cite{tu2019extreme} stations equipped with multiple modules. These high-speed charging stations could deliver up to 350 kW of rated power, enabling rapid charging cycles within 10 minutes timeframe, thereby establishing a comparable refueling experience to that of internal combustion engine vehicles (ICEVs).  

\begin{figure*}[h]
    \centering
    \includegraphics[scale=0.23]{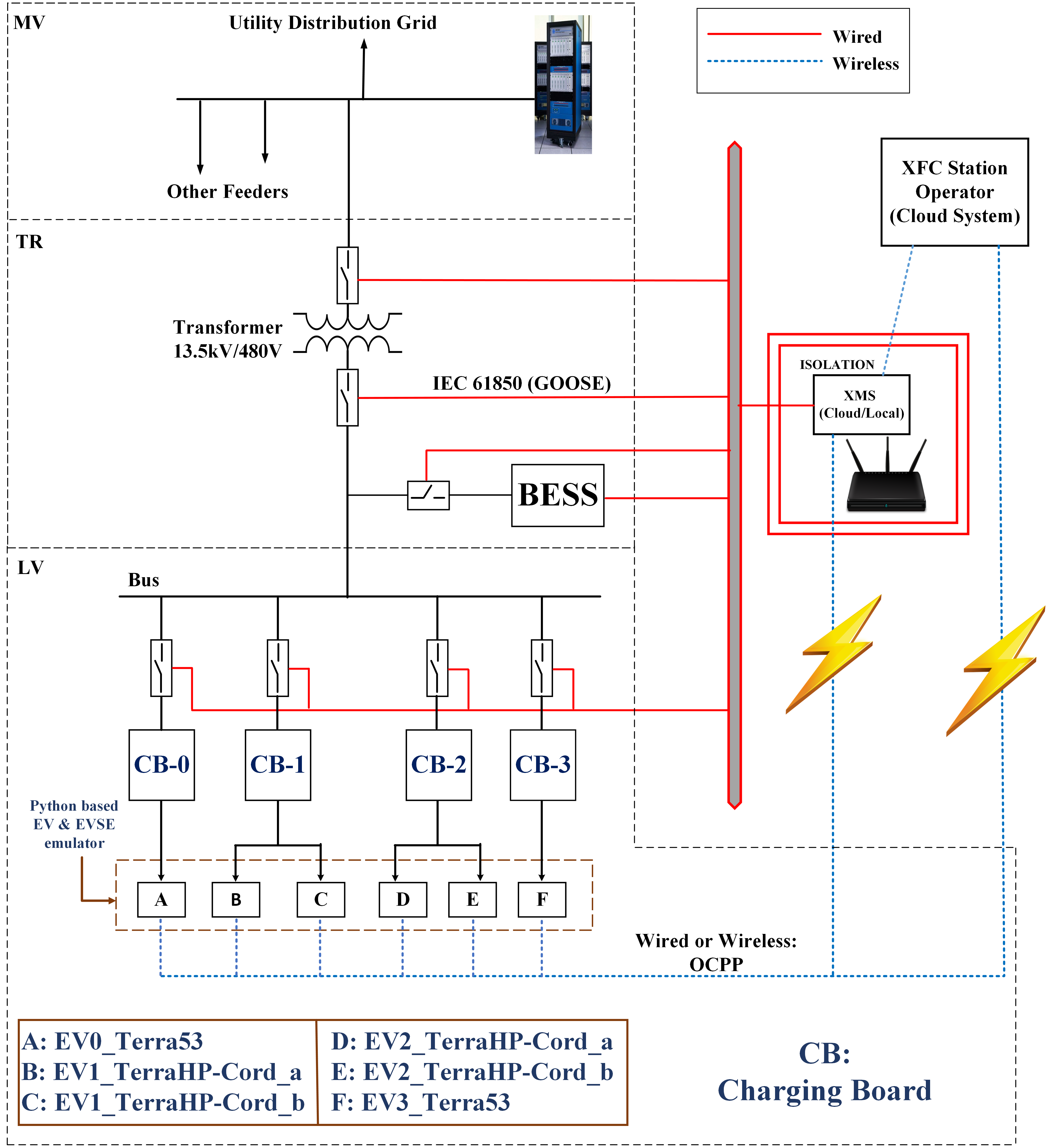}
    \caption{Proposed architecture of EVCI}
    \label{fig: EVCI as CPS}
\end{figure*}

As a consequence of rapid fast charging, the grid perceives the EVCI aggregated with EVs as a bulk power system load (during G2V mode) with inherent energy storage capacity such as BESS. When all EVs are simultaneously charged, the grid experiences various voltage and frequency stress, particularly during peak demand periods. These conditions necessitate load shedding actions, thereby compromising the overall reliability of the grid \cite{chen2021multimicrogrid}. To sustain the balance between load and generation, the EVCI actively participates in demand response (DR), entailing the charging of the EVs during the off-peak periods, and the charged EVs along with the BESS function as sources during the peak periods \cite{barati2018managing, chakraborty2024planning}. The integration of the EVCI with functions other than charging, such as load balancing, demand response, frequency response, etc., requires a communication network with internet-of-things (IoT) between various operators like the distribution system operator (DSO), EVCI administrators, and the EV users. These communication requirements across different elements of the EVCI play a significant role in defining it as a cyber physical power system (CPPS) \cite{yohanandhan2020cyber}. The CPPS is a physically interconnected system merged with cyber components encompassing computing, control, and communication functions allowing bi-directional flow of power and information thereby enabling the implementation of advanced smart grid technologies \cite{abdelmalak2022survey}. 

Due to the substantial dependence of the CPPS on the cyber systems, their communication capabilities are highly vulnerable to malicious cyber intrusions \cite{lu2022resilient}. The vulnerabilities in the EVCI \cite{acharya2020cybersecurity} may be internal and external vulnerabilities such as EV-X or EVCS-X interface vulnerabilities. The former consists of CAN (control area network) bus \cite{koscher2010experimental}, tyre pressure monitoring system (TPMS) vulnerabilities \cite{rouf2010security}. In the latter one, the X indicates various physically accessible ports, Internet Service Protocols (ISP), vendors, roadside infrastructures, and vehicular vulnerabilities. Adversaries exploit these vulnerabilities, to infiltrate the system thereby initiating a cyber attack. These attacks result in anomalies within the data, where the data exchange occurs between the diverse physical elements by way of the cyber components. The anomalies in the data of EVCI due to the cyber attacks cause service disruptions, charging delays, financial loss, safety risks, resource drain, grid instability, reputation damage, etc., \cite{acharya2020public, soykan2021disrupting, gumrukcu2022impact}. So, these anomalies need to be detected and a proper defense mechanism has to be implemented to mitigate or to prevent the effects of cyber-attacks.

%Due to these vulnerabilities, various attackers intrude on the system in various domains

%The overall srtucture of EVCI consists of the EV supply equipment (EVSE), CSMS (Charging station management systems),  Communications play a major role in the EVCI, at various stages

%Communication required for these DSM and interconnected layers make CPS, talk about protecting CPS from attacks and EVs-scope of the attack occurrence and y we are doing it

\section{related work}
\label{Section: related work}
Many researchers have addressed the cybersecurity of power grids, but the cybersecurity of EVCI has not gained much attention. The ways of cyberattacks, targeting the vital infrastructure are changing with diversified patterns in the energy distribution sector through remote communications, virtual private network (VPN) links, and corporate networks. The physical parts of the CPPS could be damaged by breaching its information and communication technology (ICT) infrastructure and obtaining more precise access to the supervisory control and data acquisition (SCADA) system which controls and monitors the components of the power grid without the need of physical attack \cite{ten2008vulnerability}. After the cyberattacks on the Ukrainian power grid, particularly in the years 2015 and 2016 showed the necessity of cyber-physical security for substations and SCADA systems, U.S. officials from the Department of Energy (DoE), the Department of Homeland Security (DHS), the Federal Bureau of Investigation (FBI), and the North American Electric Reliability Corporation (NERC) increased their efforts realizing it, as their opportunity to examine the strategies and tactics used by the aggressor to predict the likelihood occurrence of the cyberattacks and it,s type in future \cite{case2016analysis, simons2020hybrid}. In both years, the attacker's motive is to successfully compromise the industrial control system (ICS) causing potential infrastructure threats, equipment failure, and power outages. Therefore, to improve the resilience, reliability, safety, and, security of power systems, it is essential to strengthen the cybersecurity of industrial automation and control systems in the power grid.

In \cite{acharya2020cybersecurity}, various vulnerabilities are addressed that arise due to the growth of EVs from the power grid perspective. It characterizes the vulnerabilities susceptible to exploitation, imposing potential risks to EVs and EVCI equipment, the power grid, or a combination of both. The cyber security issues associated with electric chargers and the potential consequences of cyber attacks are examined in \cite{johnson2022review}. It extensively explores the conceivable cyber attacks on utility operations, power systems, interconnected systems, and billing procedures. Additionally, it deliberates on recommendations and optimal strategies for enhancing cybersecurity within EVCI systems. The attacks on EV charging stations and their users were simulated demonstrating practical impacts on the power grid leading to service disruptions and grid failures are presented in \cite{nasr2022power}. The findings emphasize the need for mitigation strategies and uncover concerns regarding the insufficient security measures implemented in deployed charging stations.

% \begin{figure*}
%     \centering
%     \includegraphics[scale=0.25]{system_architecture.png}
%     \caption{Physical architecture of EVCI}
%     \label{fig: physical architecture}
% \end{figure*}

\section{Contributions and Paper Organization}
\label{Section: Contributions and Paper Organization}

The primary contribution of this article is detecting the anomalies in the $\Delta SoC$ values of EVs when connected to the charging stations. Since there is collinearity between the SoC and the charging current values \cite{lee2023enhanced}, the charging current behavior is reflected in the SoC being a relative parameter thereby, $\Delta SoC$. A ridge regression based ML model is used for predicting the $\Delta SoC$ values. With the help of the predicted $\Delta SoC$ values, the EVCI administrator is able to detect the type of vehicle, calculate the charging time, and also in estimating the battery capacities. This work predicts the $\Delta SoC$ values from the obtained EVCI data and does not rely on the EV data. So, if there is any communication failure or incorrect estimations in the SoC of the EVs, the model won't be affected. The proposed ML model considers the current values of utility, storage system, and charging ports as the inputs of the model and $\Delta SoC$ values as the output of the model. Therefore, predicting the $\Delta$\textit{SoC} values with the current values at various ports and sensors data and the ML models helps in finding the charging behaviors of the EV, the type of the EVs arriving at the charging station, and preventing energy theft by estimating the amount of energy transferred to the EV battery with the SoC values given the battery capacity is known. 

The predicted SoC ($\Delta SoC_{predicted}$) values are compared with the EV communicated SoC values ($\Delta  SoC_{actual}$) which are assumed to be spoofed later. The absolute difference is calculated between these two data points and compared with a predefined threshold iteratively. Furthermore, the overall contributions of this article are as follows:
\begin{itemize}
    \item A grid connected EVCI system is proposed and simulated to obtain the data.
    \item The proposed EVCI model consists of both cyber and physical components including both wired and wireless communications by utilizing various protocols such as OCPP and IEC 61850 (GOOSE).
    \item A ridge regression based ML model is implemented to predict the SoC value of the EV's battery.
    \item This work detects the anomalies in the $\Delta SoC$ of the EV's battery using the EVCI data. 
    \item The anomaly is detected based on the absolute differences between the communicated and predicted $\Delta SoC$ and by observing the nature of these differences continuously.
    \item Various types of spoofing scenarios are simulated on the testing data for producing the anomaly data.
    \item In addition to anomaly detection, the proposed methodology also considers the spoofing classification.
    \item The proposed methodology is tested and validated for different cases.
\end{itemize}

This article structure commences with an introduction to the CPPS, EVCI infrastructure types, and their basic operation in Section \ref{section: Introduction}. The comprehensive review of literature on the cybersecurity aspects of EVCI when interfaced with the utility grid is presented in Section \ref{Section: related work}. Then Section \ref{Section: Contributions and Paper Organization} describes the author's contribution to the work. The subsequent Section \ref{Section: Modelling of the System} outlines the modeling framework for the proposed EVCI system. Section \ref{Section: Data acquisition and preprocessing} describes the parameters of the data acquired for both the training and testing data. It also provides the pre-processing of the data for using it as an input to the ML model. Within Section \ref{Section: Methodology}, the methodology and ML model employed for the prediction of $\Delta SoC$ are explained. The metrics used for evaluating the model performance are explained in Section \ref{Section: Case Studies}. Additionally, various case studies are also provided to validate the proposed anomaly detection framework and spoofing classification. Finally, Section \ref{Section: Conclusion} concludes with the achievements of this study while mentioning the potential areas for future research.

\section{Modelling of the System}
\label{Section: Modelling of the System}
The modeling of the proposed EVCI system is categorized into two parts: 1. Physical system modeling (representing the physical components like the charging board (CB), utility grid, BESS, etc.,) and 2. Cyber system modeling (consisting of communications, and protocols like OCPP). Fig. \ref{fig: EVCI as CPS} illustrates the comprehensive architecture of the EVCI comprising these two systems.

% \begin{figure}[h]
%     \centering
%     \includegraphics[scale=0.1]{XFC_station_model.png}
%     \caption{Charging station depot model with BESS}
%     \label{fig: XFC station model}
% \end{figure}

\subsection{Physical System Modelling}
The physical system modeling of an EVCI comprises multiple units pertaining to the power flow and energy storage. The key components in the physical system of the EVCI include the point of common coupling (PCC) linking the EVCI to the utility grid through a transformer, BESS, CBs, and charging ports. Fig. \ref{fig: EVCI as CPS} illustrates the architectural components within the physical system of EVCI, responsible for storing the root mean square (RMS) values of the various measured electrical parameters such as active power (P), reactive power (Q), voltage (V), and current (I).

The proposed system utilizes a step-down type transformer of rating 13.5kV$/$480V placed between the PCC and the EVCI as shown in Fig. \ref{fig: EVCI as CPS}. The main function of this transformer is to step down the grid level medium voltage (MV) to the low voltage (LV).  The BESS and CBs are connected to the PCC through this transformer. This article considers four CBs namely CB-0, CB-1, CB-2, and CB-3. To facilitate EV charging, the charging ports are connected to the CBs. The detailed representation of these charging ports and their current notations are provided in Table \ref{tab: representation table}. Two types of charging ports are considered in this work namely, Terra HP charger that supports from 175-350 kW fast charging and Terra 53 charger supporting up to 50 kW power \cite{ABBcharger}.

\begin{table}[h]
\centering
        \caption{Charger currents representation}
        \label{tab: representation table}
        \begin{tabular}{ccc}
            \toprule
            \textbf{Charging board}  & \textbf{Charging ports} & \textbf{Notation}\\
            \midrule
            CB-0 & EV0\_Terra53 & I$_{EV0}$\\
            \midrule
            \multirow{2}{*}{CB-1} & \multicolumn{1}{c}{EV1\_TerraHP-Cord\_a} & \multicolumn{1}{c}{I$_{EV1a}$} \\
                                 & \multicolumn{1}{c}{EV1\_TerraHP-Cord\_b} & \multicolumn{1}{c}{I$_{EV1b}$} \\
            \midrule 
            \multirow{2}{*}{CB-2} & \multicolumn{1}{c}{EV2\_TerraHP-Cord\_a} & \multicolumn{1}{c}{I$_{EV2a}$} \\
                                 & \multicolumn{1}{c}{EV2\_TerraHP-Cord\_b} & \multicolumn{1}{c}{I$_{EV2b}$} \\
            \midrule
            CB-3 & EV3\_TerraHP & I$_{EV3}$\\
            \bottomrule
        \end{tabular}
\end{table}

% s is just a common point for all the indications of the various
 
Since the main aim of this work is detecting the anomalies, the modeling of EVs is done considering only two types of vehicles whose battery types are BEV 300 and BEV 150 where BEV stands for battery electric vehicle. The BEV 300 battery type EVs are charged with a Terra HP charger that can charge up to 300 kW and the BEV 150  battery type EVs are connected to Terra 53 chargers for charging the power up to 300 kW. For all the other types of EVs, the nominal charging power is considered 50 kW and connected with Terra 53 chargers. 

It is crucial to note that any vehicle upon connecting to the EVCS transmits some amount of data. This data encompasses details such as charging protocol, EV type, arriving time, departure time, initial SoC, and target SoC \cite{chen2017dynamic}. The target SoC signifies the user-demanded SoC, representing the desired charge level the EV should attain before the scheduled departure time. The EV also continuously transmits the SoC values to the EVCI through chargers. The EVCI charges the EVs by supplying the grid power or the BESS power depending on availability.

\subsubsection{BESS modeling}
A simple model of the XFC station depot connected with a BESS is represented in Fig. \ref{fig: EVCI as CPS}. The parameters and the specifications of the BESS used in this work are provided in Table \ref{tab: BESS parameters and specifications}. This paper focuses on the modeling of public EVCS, with the assumption of uncontrollable arrival rates of EVs and unpredictable power flow to the EVSE. Since the grid power is also not controllable, Therefore the only power (or current) in the EVCI is the BESS.
\begin{table}[h]
    \centering
    \caption{BESS parameters and specifications}
    \label{tab: BESS parameters and specifications}
    \begin{tabular}{>{\centering\arraybackslash}p{4.5cm}>{\centering\arraybackslash}p{1cm}}
    \toprule
    \textbf{Parameter} & \textbf{Value}\\
    \midrule
    Maximum charging power (kW) & 500\\
    \midrule
    Maximum Discharging power (kW) & 500\\
    \midrule
    Energy capacity (kWh) & 250\\
    \midrule
    Charging efficiency (\%) & 95\\
    \midrule
    Discharging efficiency (\%) & 95\\
    \midrule
    Maximum SoC (\%) & 90\\
    \midrule
    Minimum SoC (\%) & 20\\
    \midrule
    Initial SoC (\%) & 50\\
    \midrule
    Maximum power import (kW) & 1000\\
    \midrule
    Maximum power export (kW) & 1000\\
    \bottomrule  
    \end{tabular}
\end{table}

\subsection{Cyber System Modelling}

The cyber system consists of XMS (extreme fast charging managing system) operator and XMS cloud. Both the wired and wireless communications are employed in this system. The EVSE, BESS, and transformer parameter measurement circuits are communicated to the XMS cloud in a wired mode through the IEC 61850 messaging protocol, and the charging ports and the EVs communicate to the XMS operator and the XMS cloud system in a wireless mode through OCPP (Open Charge Point Protocol). Fig. \ref{fig: EVCI as CPS} shows different modes of communication, connecting the various parts of the proposed system.

The IEC 61850 is a standard messaging protocol used for multicast messages such as generic object-oriented substation events (GOOSE). It recommends only message integrity and authenticity. Generally, in substation automation systems, GOOSE messages are used to carry the breaker open or close commands \cite{hussain2020method}. Another protocol used here is OCPP, which is an open source to facilitate the communication between the charging ports and the backend systems. This was proposed by the Dutch foundation ElaadNL and should be standardized and certified by OCA (Open Charge Alliance) to ensure uniformity and cross-vendor compatibility \cite{garofalaki2022electric}. The electrical parameters data is transmitted from the physical parts of the system through these protocols. Besides, the charging ports also transmit another parameter known as charge status (CS) which is a binary that returns 1 when the EV is charging and 0 otherwise. Both the physical and cyber systems simulations are performed in real-time on the OPAL-RT simulation platform using both MATLAB and Python environments.

\begin{figure}[h]
    \centering
    \includegraphics[scale=0.15]{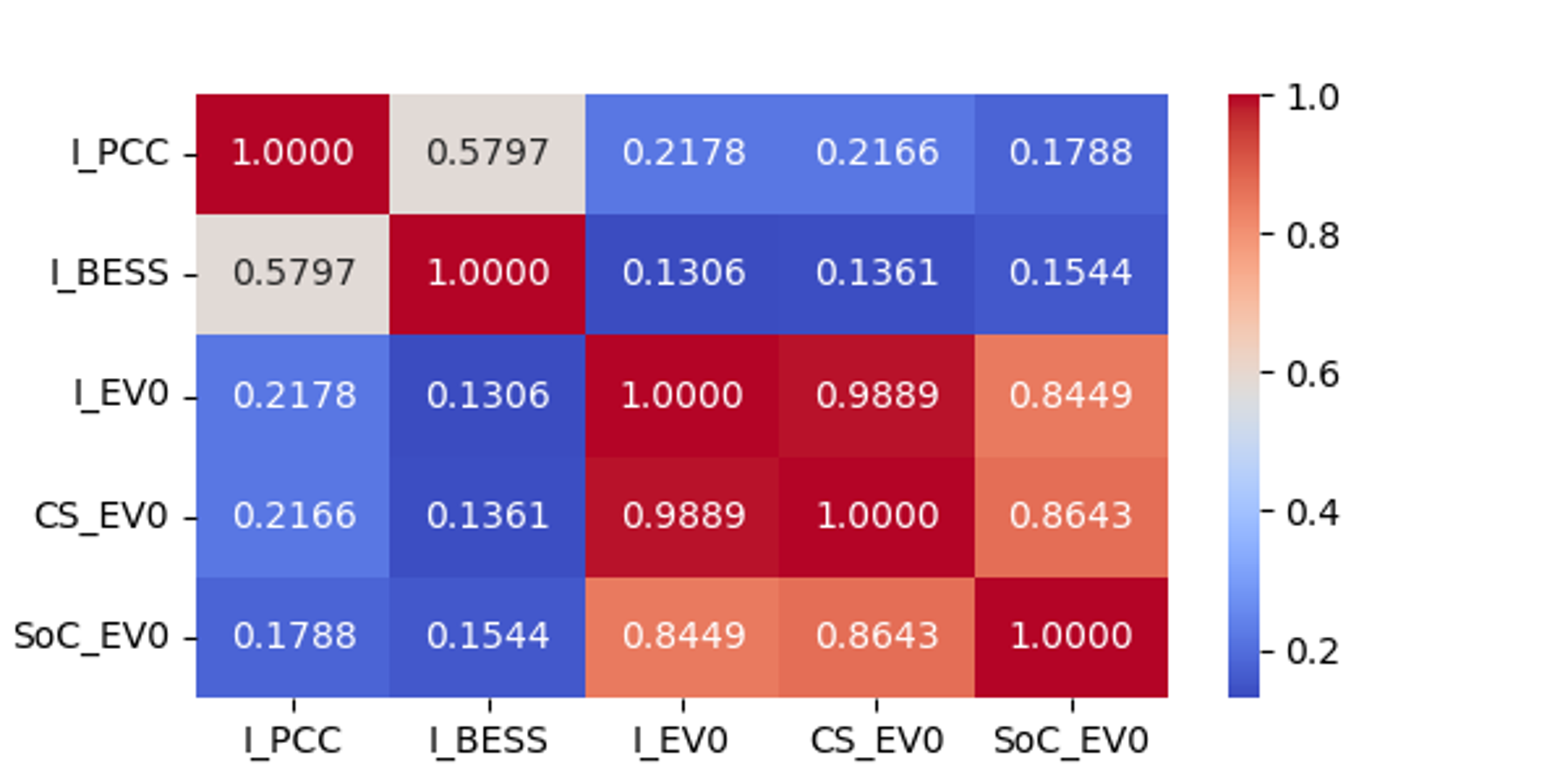}
    \caption{Heatmap of correlation among various parameters related to charging port EV0$\_$Terra53}
    \label{fig: correlation heatmap}
\end{figure}

\section{Data acquisition and preprocessing}
\label{Section: Data acquisition and preprocessing}
The data required for developing the proposed methodology is acquired and stored for both training and testing separately. The training data is collected for four days at a rate of one sample per second and the testing data is collected for one day at the same rate. The acquired data consists of variables PPC current (I$_{PCC}$), BESS current (I$_{BESS}$), charging port current (I$_{EVn}$), active power (P$_{EVn}$) and reactive power (Q$_{EVn}$) of charging port, charge status ($CS$)  where $n$ is the representation of the EV charging port. Since it is required for the EVCI to know the SoC of the connected EV, The SoC of the connected EV is continuously transmitted to the EVCI through OCPP as explained later. This data is also collected and stored in the variable SoC$_{EVn}$ at an equivalent sampling rate.

Since the main aim of this article is detecting the anomalies in $\Delta SoC$, it is calculated by subtracting the present instant SoC value from its previous instant. This $\Delta SoC$ calculation is performed for all the EVs connected to various charging ports and stored in variables labeled as $\Delta SoC_{EVn}$. It is the actual or calculated SoC, termed as $\Delta SoC_{actual}$. This data variable is the output of the proposed ML model where various types of spoofing techniques are assumed to be done in the form of a cyber attack. 
This study explores three distinct types of spoofing techniques: decimal shifting, incremental array spoofing, and random spoofing. Table \ref{tab: spoofing table} presents the numerical levels associated with each type of spoofing technique. These values are chosen to simulate stealthy attack scenarios \cite{sui2020vulnerability} and implemented individually on the testing data at different instances.

\begin{table}[]
    \centering
    \caption{Spoofing techniques: value ranges and types}
    \label{tab: spoofing table}
    \begin{tabular}{p{2.5cm}p{5cm}}
    \toprule
        \textbf{Noise type} & \textbf{Range (or) Type}\\
        \midrule
        Random & Uniform distribution (in range of 10$^{-2}$)\\
        \midrule
        % Scaling by decimals & 0.88 to 0.99\\
        % \midrule
        Shifting in decimals & $-0.009$ to $+0.009$\\
        \midrule
        Incremental array & Arithmetic progression\\
        \bottomrule
    \end{tabular}
\end{table}

Fig. \ref{fig: actual vs predicted} shows the communicated and predicted $\Delta SoC$ values where the spikes in communicated $\Delta SoC$ values signify the large variation in it that occurs due to the arrival or departure of the EVs at the respective charging port of the EVCI. Since these data points are not considered anomalies, the spoofing techniques are simulated to the particular windows of the obtained time series data with lengths more than the length of these spike durations. The random spoofing replaces the $\Delta SoC_{actual}$ values with random values within the specified range by following a uniform distribution. Similarly, the incremental array spoofing technique also replaces a series of $\Delta  SoC_{actual}$ values, with these series of values should be in arithmetic progression. The scaling and shifting in decimals type of spoofing is performing mathematical calculations to the actual output values for simulating the spoofing scenario. The actual output of the time series data is replaced with the spoofed time series output of the testing data and later compared with the output of the ML model ie., $\Delta  SoC_{predicted}$ for anomaly detection.

For any ML model, the input data has to be preprocessed. Due to the spikes and high variations in $\Delta SoC$ values, the min-max normalization fails because of the irregular distribution of data. The standard scaling method is used to preprocess the data as required by the Ml model.

\section{Methodology}
\label{Section: Methodology}
\subsection{Multicollinearity}
Multicollinearity is defined as the phenomenon of the existence of a strong correlation between various variables in a regression model \cite{ridgearticle}. This phenomenon arises from diverse circumstances, including the inherent physical interpretations of variables, such as metrics or measurements utilized within software defect prediction models. The interdependence among these variables could be influenced by the intrinsic nature of their meanings, contributing to the multicollinearity. Another reason for multicollinearity is the large collection of metrics. For constructing a better model we tend to collect metrics fitting to the defects as many as possible resulting in serious multicollinearity issues in datasets \cite{yang2018ridge}.

The multicollinearity in the acquired datasets exists between the battery current of the EV (I$_{EVn}$) and its estimated SoC values, where n indicates the charging port to which the EV is connected. Since coulomb counting \cite{ng2009enhanced} is the most adopted method, it estimates the SoC as a relative measure of charging or discharging EV current and integrates it over time and is given as follows:
\begin{equation}
    \textit{SoC}(t) = \textit{SoC}(t-1) + \frac{I(t)}{Q_{n}} \Delta \textit{t}
\end{equation}
Where $SoC(t)$ is the SoC to be estimated at present instant, $SoC(t-1)$ is the SoC of the battery at the previous time instant. $I(t)$ is current at present time instant, \textit{Q$_{n}$} is the capacity of the battery pack and $\Delta t$ is the discrete sampling period of the battery management systems (BMS). Fig. \ref{fig: correlation heatmap} demonstrates the correlations with the help of heat-map between the various variables showing the high collinearity between the current and SoC values.

There also exists another strong collinearity between the multiple input variables of the ML model such as I$_{PCC}$, I$_{BESS}$, and I$_{EVn}$ (currents flowing through all the EVs). This collinearity is easily verified by Kirchoff's current law (KCL) and is given as follows:
 \begin{equation}
     \textit{I$_{PCC}$} = \textit{I$_{BESS}$} + \sum_{k= 1}^{N}\textit{I$_{CBk}$}
 \end{equation}
Where I$_{CBk}$ is the EVCI's CB current whose values are equal to the sum of all charging ports current connected to the particular CB. N is the total number of CBs present in the charging station.

\begin{figure}[h]
    \centering
    \includegraphics[scale=0.18]{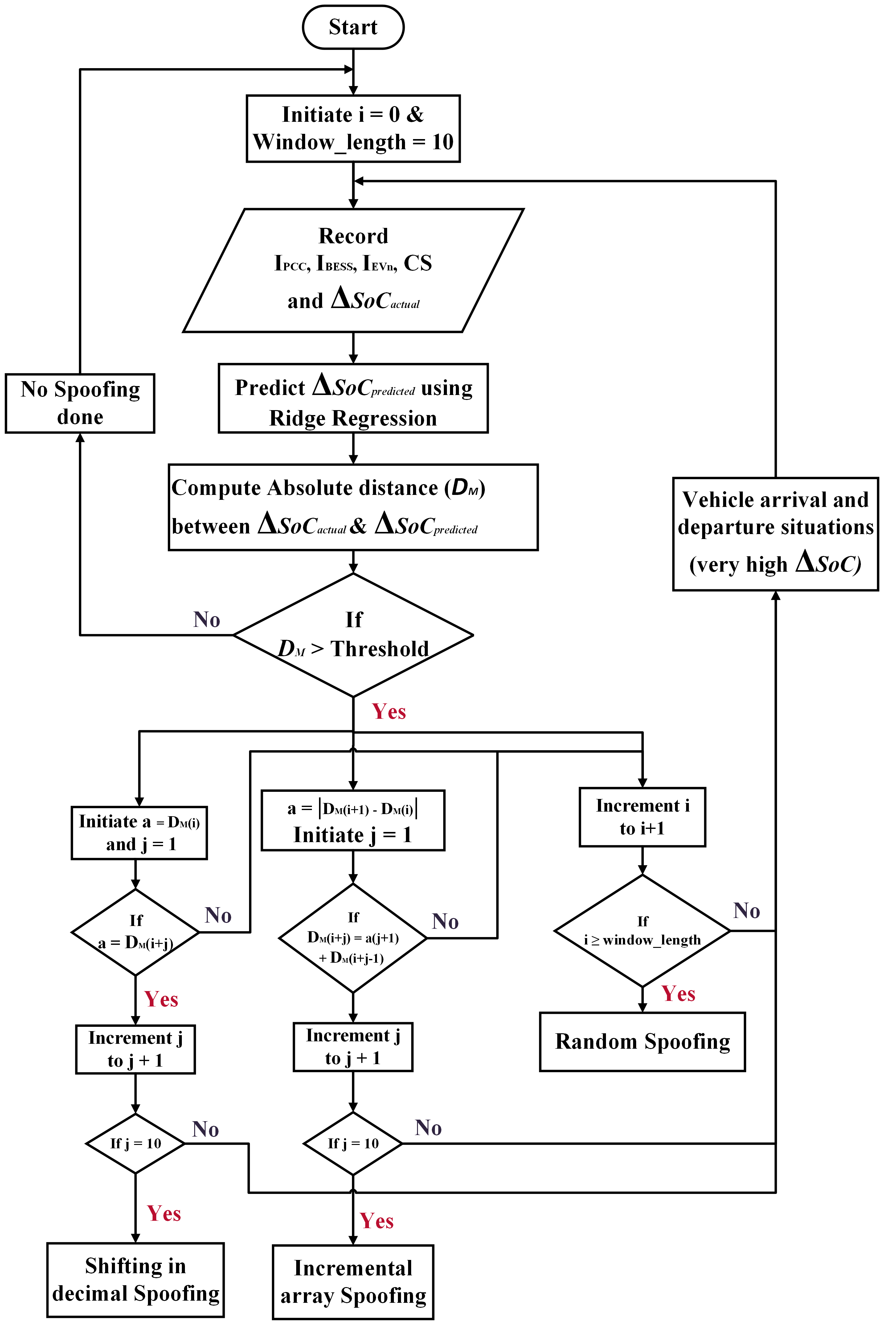}
    \caption{Proposed Anomaly detection methodology flowchart}
    \label{fig: Anomaly flowchart}
\end{figure}

\subsection{Ridge Regression}
In the context of multiple linear regression applications, ridge regression is the most commonly used parameter forecasting technique particularly used to overcome the drawbacks associated with collinearity that arise frequently in other regression scenarios \cite{mcdonald2009ridge}. It helps to mitigate the negative impact of collinearity on the least squares (LS) estimator. The main feature of this ridge regression is addressing the problem of collinearity without removing any data variable. With the other regression estimators, there is a probability of coefficients becoming larger in absolute values and sometimes may also have a wrong sign. The likelihood of encountering these challenges increases the deviation in prediction vectors from orthogonality. As discussed in the previous sections, due to the collinearity present between the multiple input variables in this work, considering the  multiple linear regression standard model:

\begin{equation}
    \textbf{y} = \textbf{x}\beta + \varepsilon
\end{equation}
Where $\mathit{E}(\varepsilon) = 0$, $\mathit{E}(\varepsilon \varepsilon')=\sigma^{2}I_{n}$, and \textbf{x} is a full rank matrix of n$\times$p dimensions. The bold letter symbols denote vectors and matrices. It is assumed that the variables are standardized to the correlation form (\textbf{x}'\textbf{x}) and the response variable with each of the explanatory variables is given by the vector $\gamma$ $\equiv$ \textbf{x}'\textbf{y}. Let $\hat{\beta}$ be the LS estimate of the variable $\beta$ and is given by

\begin{equation}
    \hat{\beta} = (\textbf{x}'\textbf{x})^{-1}\textbf{x}^{-1}\textbf{y}
\end{equation}

The average distance between the $\beta$ and $\hat{\beta}$ directly contributes to the challenges in this standard estimation. It is achieved by the standard squared distance (\textit{L}$^{2}$) between $\beta$ and $\hat{\beta}$ with the following properties:

\begin{eqnarray}
    L^{2} = (\hat{\beta} - \beta)'(\hat{\beta} - \beta)\\
    \mathit{E}(L^{2}) = \sigma^{2} \text{trace} (\textbf{x}'\textbf{x})^{-1}\\
    \mathit{E}(\hat{\beta}'\hat{\beta}) = \beta'\beta + \sigma^{2} \text{trace} (\textbf{x}'\textbf{x})^{-1}
\end{eqnarray}

\begin{figure}[]
    \centering
    \includegraphics[scale=0.3]{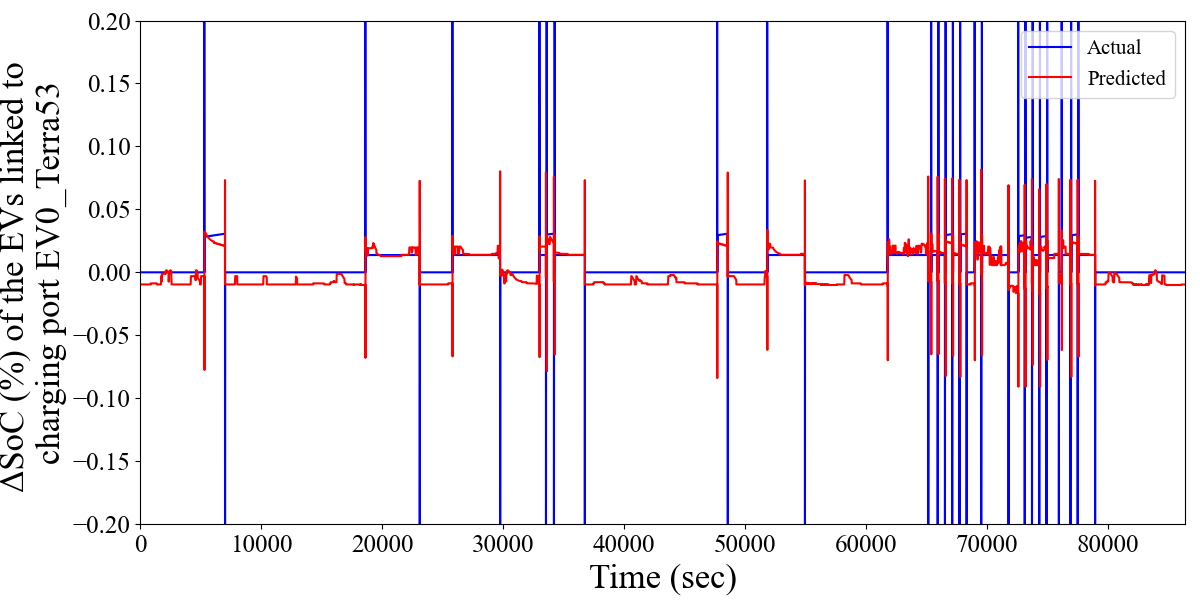}
    \caption{Comparison of actual and predicted $\Delta SoC$ values of the EVs connected to charging port EV0$\_$Terra53}
    \label{fig: actual vs predicted}
\end{figure}

% For a given n training variables x$_{i}$ = (x$_{i,1}$, x$_{i,2}$,......x$_{i,d}$), $\forall$ i:1 to n, x$_{i}$ is the ith parameter vector data variable and  x$_{i,j}$ is the j$^{th}$ metric values of the vector x$_{i}$, and d is the total number of metrics. When dealing with extremely collinear data, traditional linear regression suffers difficulties with coefficient stability since the design matrix (x) approaches singularity due to correlated independent variables. Ridge regression is a realistic solution that penalizes excessive coefficients and promotes their shrinking toward zero by incorporating an L2 penalty term into the loss function (λ ∥β∥22). The addition of the L2 penalty term encourages smaller and more stable coefficient estimates. The parameter (λ, α) governs the trade-off between fitting the data well (first term) and keeping the coefficients small (second term). The ridge regression solution is given by

Traditional linear regression methods, when dealing with high collinear data suffer from coefficient stability as the input matrix (\textbf{x}) becomes singular due to correlated independent variables. This paper adopted ridge regression to provide a realistic solution by penalizing excessive coefficients shrinking toward null values, by incorporating the L2 normalization penalty term into the loss function ($\lambda\left\| \beta \right\|^{2}_{2}$). This penalty term makes the model coefficients smaller and more stable. The parameters $\lambda$ and $\alpha$ govern the trade-off between achieving a good fit of the model and restricting the coefficients to small values respectively. The final solution for the ridge regression is given as follows:
\begin{equation}
\hat{\beta}_{ridge} = \text{argmin}_{\beta}\left\{ \left\| y - X\beta \right\|^{2}_{2} + \alpha\left\| \beta \right\|^{2}_{2} \right\}
\end{equation}

Grid search and cross-validation (GS-CV) method is employed to rigorously evaluate a set of variables \cite{GS-CV}. This involves assessing the model's performance across each fold of cross-validation. To ensure the model's generalization ability beyond the training data, the optimal configuration was selected based on the fold-averaged performance metric. The choice of specific parameters influences the extent of coefficient shrinkage subsequently impacting the stability and accuracy of the model. Through GS-CV, the value of the parameter achieves a good trade-off between variance reduction and minimum bias \cite{adnan2022utilizing}. 

% We employed a method combining grid search and cross-validation (GS-CV) to rigorously evaluate a set of variables within a predetermined grid. This involved assessing the model's performance (mean squared error) across each fold of cross-validation. The optimal configuration was then selected based on the fold-averaged performance metric, ensuring the model's ability to generalize beyond the training data. Notably, the choice of a specific parameter significantly influenced the extent of coefficient shrinkage and consequently impacted the model's stability and accuracy. Through GS-CV, we identified the parameter value that minimized cross-validated error, achieving a delicate balance between reducing variance and introducing minimal bias. 

% The Ridge Regression model was implemented using the Scikit-learn library, and the grid search was conducted with a k-fold cross-validation strategy $(k=20)$ to ensure robust model assessment. The performance of each candidate model was evaluated using the negative mean squared error as the scoring metric.The optimal hyperparameters were determined to be $\alpha=10.0$ and solver set to '$svd$'.

This basic strategy reduces multicollinearity by providing three critical benefits: reduced variance (ensuring stable estimates), improved bias-variance trade-off (favoring generalizability), and higher model stability (reducing noise sensitivity). As a result, ridge regression enables researchers to build robust and interpretable models even when confronted with tough collinear data, improving the quality of scientific discourse and encouraging more robust conclusions.

The performance of the six regression models for all the six EV charging ports is evaluated individually with suitable metrics. Table \ref{tab: metrics table} depicts the various metrics such as mean squared error (MSE) and its values are also provided to evaluate the model.

% \subsection{Mahalanobis distance}
% It is a measure of the distance used to determine the distance between a point and a distribution. Unlike the Euclidean distance (which measures the straight line distance between two points in an n-dimensional space by considering all the dimensions equally). The Mahalanabolis distance correlations between dimensions and different variances along each dimension. It has more applications in the case of multivariate data with interrelated variables having different scales.

% The Mahalanobis distance between a point (X) and the distribution with mean vector ($\mu$) and a covariance matrix ($\Sigma$) is calculated as follows:
% \begin{equation}
%     D_{M} = \sqrt{(X - \mu)^{T} \Sigma^{-1} (X - \mu)}
% \end{equation}
% Where X represents the data point for which the distance has to be calculated, $\mu$ is the mean vector of the distribution (average values of each variable in the dataset). $\Sigma$ is the covariance matrix and it represents the relationships between different variables and their variances. If these variables are not related to each other, the covariance matrix is an identity matrix indicating that all variances are 1 and covariances are 0. In this case, the Mahalanobis distance reduces to the Euclidean distance.

\begin{algorithm}
	\caption{Anomaly Detection}
        \label{Algorithm-1}
	\begin{algorithmic}
  	\State Initialize the value threshold and window length as max$\_$iter.
        \State Record the values $I_{PCC}$, $I_{BESS}$, $I_{EVX}$, $CS$, and  $\Delta SoC_{actual}$.
        \For {$i=1,2,\ldots$max$\_$iter}
        \State Predict $\Delta SoC_{predicted}$ using ridge regression ML model.
        \State Compute the absolute difference (D$_{M}$) between the values $\Delta SoC_{actual}$ and $\Delta SoC_{predicted}$.
        \If{D$_M$ $\geq$ threshold}
        \State x $ \gets D_M(i)$, y $\gets \mid D_M(i+1) - D_M(i)\mid$. 
        \State Go to Algorithm-\ref{Algorithm-2}, Algorithm-\ref{Algorithm-3}, and Algorithm-\ref{Algorithm-4}.
        \EndIf
		\EndFor
	\end{algorithmic} 
\end{algorithm}

The comparison between actual and predicted $\Delta SoC$ values of the EVs connected to the charging port EV0$\_$Terra53 are shown in Fig. \ref{fig: actual vs predicted}. Since there are large deviations in the actual $\Delta SoC$ values, these scenarios occur during the arrival and departure of the EVs to the EVCI as depicted in Fig. \ref{fig: actual vs predicted}, it is due to the reason that during the EV arrival to the EVCI, SoC changes abruptly from 0 to a greater value and also during departure, the SoC drops from a higher value to 0. So, the proposed methodology considers a time window of different lengths for the different charging ports. As a result, whenever the D$_{M}$ exceeds the threshold value, the methodology starts the iteration count and checks for the larger thresholds in the consecutive data samples throughout the window. If the D$_{M}$ exceeds the threshold throughout the window then the methodology detects it as an anomaly. The Algorithm \ref{Algorithm-1} provides the pseudocode for the anomaly detection.

\begin{algorithm}
	\caption{Shifting in Decimals Spoofing}
        \label{Algorithm-2}
	\begin{algorithmic}[1]
        \State $a\gets D_M(i)$
        \For {$j=1,2,\ldots$max$\_$iter}
        \If{$a = D_M(i)$}
		\State $j \gets j+1$
        \If{j = max$\_$iter}
        \State Display \textit{Shifting in decimals Spoofing detected}
        \
        \EndIf
        \EndIf
        \EndFor
        \end{algorithmic} 
\end{algorithm}

\begin{algorithm}
	\caption{Incremental Spoofing}
        \label{Algorithm-3}
	\begin{algorithmic}[1]
		\State $a \gets \mid D_M(i+1) - D_M(i)\mid$
        \For {$j=1,2,\ldots$max$\_$iter}
        \If{$D_M(i+j) = (a*j) + D_M(i+j+1)$}
        \State $j \gets J+1$
        \If{j = max$\_$iter}
        \State Display \textit{Incremental array spoofing detected}
        \EndIf
        \EndIf
        \EndFor
	\end{algorithmic} 
\end{algorithm}

In addition to anomaly detection, the proposed methodology also categorizes the type of spoofing that causes anomaly in the target data i.e., $\Delta SoC$. Since this work considers the three types of spoofing techniques namely, shifting in decimals, incremental array spoofing, and random spoofing, the proposed framework classifies in parallel to the anomaly detection. The Algorithm \ref{Algorithm-2} describes the procedure for detecting the shifting in decimals type of spoofing which records the absolute difference once it reaches the threshold and compares it with the other succeeding values for the entire window duration if all the values are equal then the resulting spoofing is of shifting in decimals type.

The incremental array spoofing is detected by observing the series of absolute differences between the predicted and communicated $\Delta SoC$ for the window if the initial distance reaches the threshold. If any progression is observed the proposed methodology returns the type of progression present in the data. The flow diagram for this type of spoofing detection is provided in Algorithm \ref{Algorithm-3}. If the spoofing does not belong to the any of aforementioned categories, then it is termed as random spoofing. It compares the absolute differences with the threshold iteratively only if the conditions of the shifting in decimals and incremental spoofing aren't achieved as explained in Algorithm \ref{Algorithm-4}. The overall framework of the proposed methodology combining all the pseudocodes is depicted in Fig. \ref{fig: Anomaly flowchart}. Furthermore, the step-wise procedure of the proposed methodology is as follows: 

\begin{enumerate}
    \item [1] Initialize the values iterative count (i), window length, and threshold.
    % \item [2] Record the Values I$\_$PCC, I$\_$BESS, I$\_$EVX, CS, and $\Delta SoC\_act$ (communicated from the EVs).
    \item [2] Predict the output of the model ($\Delta  SoC_{predicted}$).
    \item [2] Compute D$_M$ = abs($\Delta SoC_{predicted}$ - $\Delta SoC_{actual}$).
    \item [4] if D$_M$ $\leq$ threshold go to step 2. else, go to next step.
    \item [5] x = D$_M(i)$, y = $\mid D_M(i+1) - D_M(i) \mid$ and j = 1.
    \item[6] If $D_M(i+j)$ = a, increment j to j+1\\
    else, go to step 8.
    % \item [9] If $D_M(i+j)$ = a, increment j to j+1.
    \item[7] If j = 10, display shifting in decimal spoofing detected and break.
    \item[8] If y = $\mid D_M(i+1) - D_M(i) \mid$, increment j to j+1\\
    else, go to step 10.
    \item[9] If j = 10, display Incremental array spoofing detection and break. 
    \item[10] If i $<$ 10, increment i to i+1\\
    else, display random spoofing detected.
\end{enumerate}

\begin{algorithm}
	\caption{Random Spoofing}
        \label{Algorithm-4}
	\begin{algorithmic}[1]
        \If{D$_M(i) \geq threshold$}
		\State $i \gets i+1$
        \If{i = max$\_$iter}
        \State Display \textit{Random type Spoofing detected}
        \EndIf
        \EndIf
        \end{algorithmic} 
\end{algorithm}

\section{Results \& Case Study}
\label{Section: Case Studies}
Various ML and neural network (NN) models are tested for the acquired data by preprocessing with standard scaling. The parameters and the performance metric values used in this model are shown in Table \ref{tab: metrics table}. The MSE values of the different model shows that the ridge regression model is the best fit for predicting $\Delta SoC$ values using the EVCI data. The ride regression model was implemented using scikit-learn in Python environment, and robust model assessment was ensured by conducting the grid search with k-fold cross-validation strategy with k = 20. The optimal hyperparameter $\alpha$ is determined to be 10.05. This ridge regression model is used for predicting the $\Delta SoC$, and the performance of the model is also evaluated by metrics such as mean squared error (MSE). Table \ref{tab: metrics table} depicts the values of the MSE metric of all the ML models used for each charging port. The testing data is used to validate the proposed methodology by spoofing it for various cases are shown with the obtained results. 
\begin{table}[h]
    \centering
    \caption{performance metrics comparison of various models for predicting $\Delta SoC$}
    \label{tab: model comparisons}
    \begin{tabular}{>{\centering\arraybackslash}p{2.3cm}>{\centering\arraybackslash}p{3.8cm}>{\centering\arraybackslash}p{1.3cm}}
         \toprule   
         \textbf{Model} & \textbf{Parameters} & \textbf{MSE}\\
         \midrule
         Linear regression & fit$\_$intercept = false & 1.771121117\\
         \midrule
         Multi-layer perceptron & Adam optimizer, Dropout = 0.2 & 1.77119094\\
         \midrule
         Support vector regression (SVR) & C = 1.5, epsilon = 0.18 & 1.77747376\\
         \midrule
         Random Forest (RF) & number of estimators = 230 & 2.01821433\\
         \midrule
         Ridge regression & alpha = 10 & 0.000194\\
         \bottomrule
    \end{tabular}
\end{table}

\begin{table}[h]
    \centering
    \caption{Proposed regression model performance across various charging ports}
    \label{tab: metrics table}
    \begin{tabular}{cc}
        \toprule
         % \multicolumn{1}{c}{\multirow{2}{*}{Charging Port}} & \multicolumn{2}{c}{Metrics}\\
         % \cmidrule{2-3}
         % \multicolumn{1}{c}{} & MSE & R$^2$\\
         \textbf{charging Port} & \textbf{MSE}\\
         \midrule
         EV0$\_$Terra53 & 0.000194\\
         \midrule
         EV1$\_$TerraHP-Cord$\_$a & 0.000217\\
         \midrule
         EV1$\_$TerraHP-Cord$\_$b & 0.000180\\
         \midrule
         EV2$\_$TerraHP-Cord$\_$a & 0.000324\\
         \midrule
         EV2$\_$TerraHP-Cord$\_$b & 0.000129\\
         \midrule
         EV3$\_$TerraHP & 0.000356\\
         \bottomrule
    \end{tabular}
\end{table}

% \begin{table}[h]
%     \centering
%     \caption{Performance evaluation metrics}
%     \label{tab: metrics}
%     \begin{tabular}{p{1cm}p{1.5cm}p{1.5cm}p{1.5cm}}
%     \toprule
%     \textbf{Metric} & \textbf{Random Spoofing} & \textbf{Decimal shifting} & \textbf{Incremental spoofing}\\
%     \midrule
%     \textbf{Accuracy} & 99 \% & 99.31 \% & 99.84 \% \\
%     \midrule
%     \textbf{Recall} & 0.82 & 0.87 & 0.99\\
%     \bottomrule
%     \end{tabular}
% \end{table}

The different spoofing types considered in this work are shifting by decimals, incremental array shifting, and random spoofing. These are implemented on the charging port EV0$\_$Terra53 and the proposed anomaly detection methodology is tested for all types of spoofing. since the results are shown for only one charging port the proposed framework, could be scaled to the other remaining charging ports also. 
The results obtained for detecting the anomaly caused by the different spoofing types on the charging port EV0$\_$Terra53 are further discussed as different case studies. 

\begin{figure*}[h]
    \centering
    \includegraphics[scale=0.35]{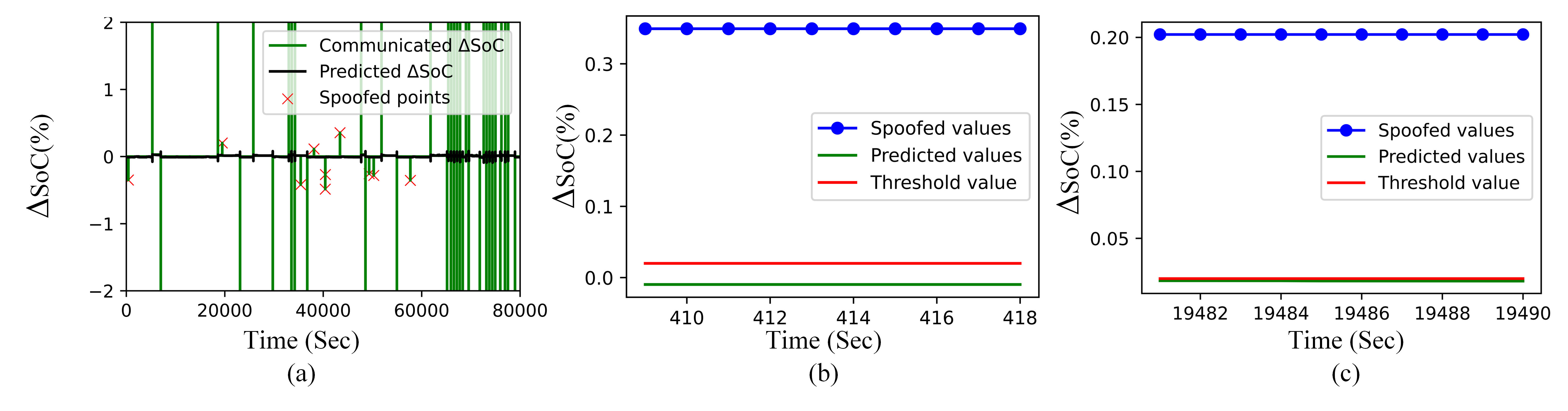}
    \caption{Shifting in decimals case study indicating (a) communicated and predicted $\Delta SoC$ values highlighting the spoofed points, anomaly detection for the spoofed windows for the instances starting with (b) 409 and (c) 19481 each window of size 10 samples}
    \label{fig: shifting in decimals}
\end{figure*}

\begin{figure*}[h]
    \centering
    \includegraphics[scale=0.35]{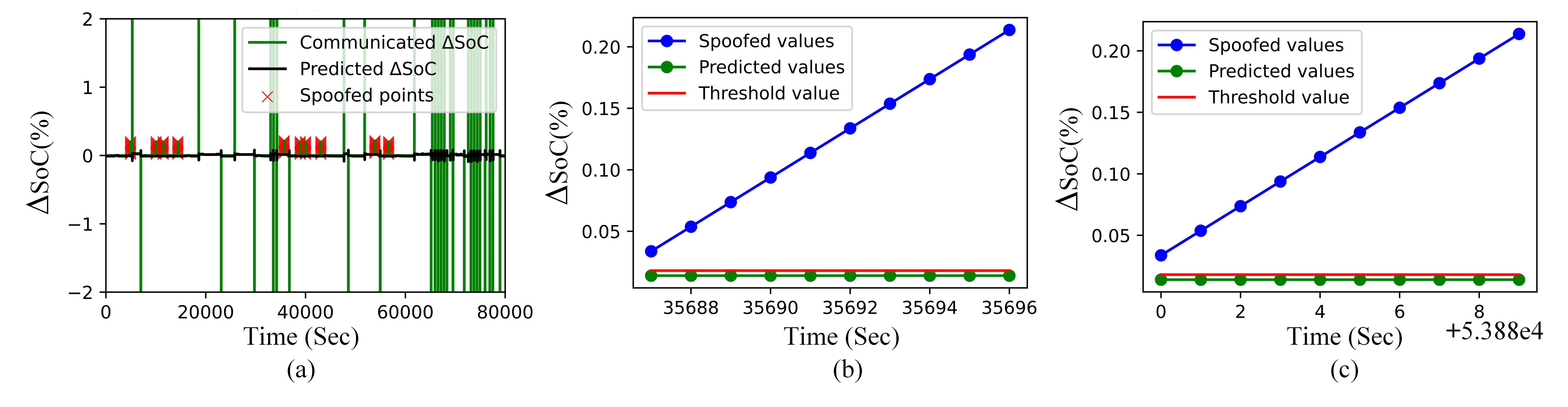}
    \caption{Incremental array shifting case study indicating (a) communicated and predicted $\Delta SoC$ values highlighting the spoofed points, anomaly detection for the spoofed windows for the instances starting with (b) 35687 and (c) 53880 each window of size 10 samples}
    \label{fig: incremental shifting}
\end{figure*}

\begin{figure*}[h]
    \centering
    \includegraphics[scale=0.35]{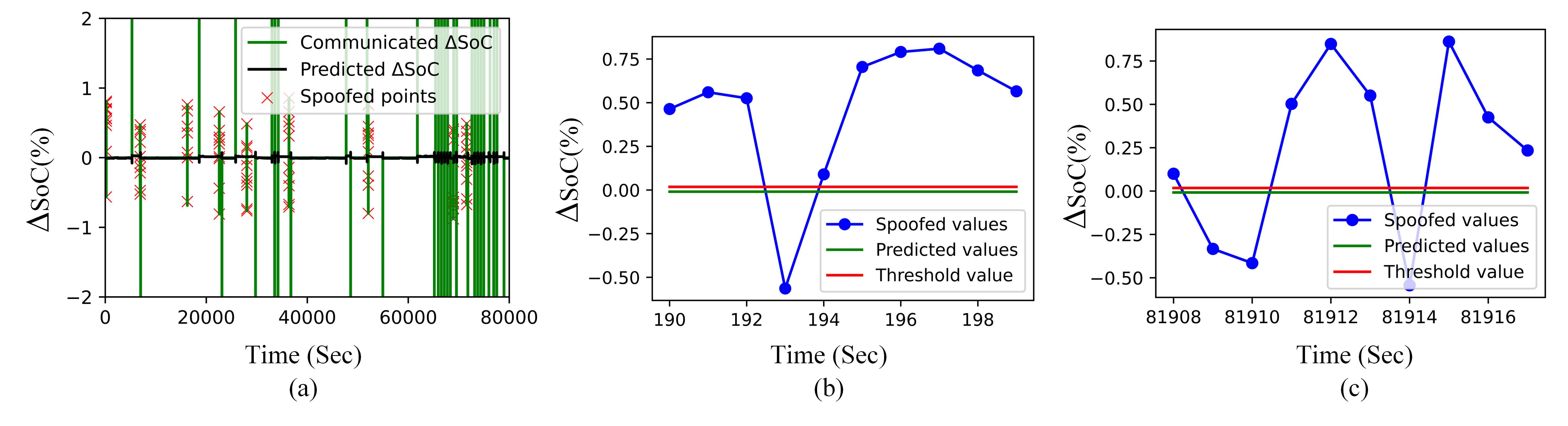}
    \caption{Random array spoofing case study indicating (a) communicated and predicted $\Delta SoC$ values highlighting the spoofed points, anomaly detection for the spoofed windows for the instances starting with (b) 190 and (c) 81908 each window of size 10 samples}
    \label{fig: random spoofing}
\end{figure*}

\subsection{Case Study: Sifting in decimals}
This case consists of shifting the actual or communicated $\Delta SoC$ values of the connected EVs. Since this work considers shifting the decimals for a smaller window size, it does not have any impact on the EVCI. However, shifting the communicated $\Delta SoC$ values continuously for longer durations may affect the EVCI system and also the EV battery. A few sample windows of size 10 samples each are considered at random instances on the testing data for spoofing. Fig. \ref{fig: shifting in decimals}(a) shows the actual communicated $\Delta SoC$ (spoofed) and the predicted $\Delta SoC$ values along with spoofed data points marked distinctly, for the charging port EV0$\_$Terra53 of the testing data. The shifting of $\Delta SoC$ values is done by adding a constant number to the communicated $\Delta SoC$ values. If this spoofing increases the actual value then EVCI assumes that EV is charged at higher speeds and the charging process is commuted without achieving the desired task. In cases of spoofing with lower values, the charging process is not commuted even after achieving the desired task. hence, these issues are prevented by detecting this type of spoofing. A threshold is predefined for this anomaly detection based on the type of EV connected to the respective charging port. Fig. \ref{fig: shifting in decimals}(b) and Fig. \ref{fig: shifting in decimals}(c) depicts the anomaly detection for this type of spoofing for different time windows at time instances 190 and 81908 respectively along with the threshold where the spoofed values are the communicated $\Delta SoC$ values. The results show that the proposed anomaly detection methodology works fine and it also classifies this shifting type spoofing with an accuracy of 99.31$\%$.

\subsection{Case Study: Incremental array shifting}

The incremental shifting of the $\Delta SoC$ consists of the shifting of the communicated $\Delta SoC$ values to a significant amount. The earlier shifting in decimal consists of adding or replacing the actual values with a constant value, whereas this shifting type adds or replaces the actual values with a series of progression values. This type of spoofing plays a crucial role during the EVs arrival and departure duration where the $\Delta SoC$ values are in higher magnitudes. The communicated (spoofed) values and the predicted values of the $\Delta SoC$ are shown in Fig. \ref{fig: incremental shifting}(a) with the spoofed data points marked distinctly for the EVs arriving at the charging port EV0$\_$Terra53 of the testing data. This spoofing makes the EVCI assume that the EV is charging at faster or slower rates than desired which results in disturbing the charging and discharging rates (C-rates) of the EVs. To mitigate these issues, this type of spoofing is detected by employing a predefined threshold. Since this progression based shifting may be incremental or decremental, the proposed methodology is verified for the incremental array shifting but it could be applied to both types of spoofing. This work considers shifting the $\Delta SoC$ values by adding a series of the arithmetic progression to the existing values thus making the SoC increase at a constant faster rate. Fig. \ref{fig: incremental shifting}(b) and Fig. \ref{fig: incremental shifting}(c) show the anomaly detection window for this type of spoofing considered at a faster rate in the testing data for the different instances. These results indicate that the proposed methodology has 90.84$\%$ accuracy in detecting the anomalies attributed to spoofing, specifically related to the array shifting of the SoC values with a sequence of arithmetic progression values.

\subsection{Case Study: Random Spoofing}

The type of spoofing techniques that do not belong to any of the aforementioned categories are classified as random spoofing. Random spoofing consists of replacing the actual$\Delta SoC$ values with random numbers. A few sample windows of size 10 instances each are spoofed and these time windows are chosen randomly. The communicated (spoofed) and the predicted $\Delta SoC$ values are shown in Fig. \ref{fig: random spoofing}(a) along with the spoofed points marked distinctly, for the charging port EV0$\_$Terra53 of the testing data. The effect of random spoofing is irregular variations in $\Delta SoC$ values resulting in the EVCI not achieving the desired tasks. This work considers replacing the actual $\Delta SoC$ value with a series of random numbers within the specified range as presented in Table \ref{tab: spoofing table}. Fig. \ref{fig: random spoofing}(b) and Fig. \ref{fig: random spoofing}(c) show the anomaly detection where the spoofing is of random type. The results show that the proposed methodology has 93$\%$ accuracy in detecting the random spoofing and classifying it.

\section{Conclusions and Future Work}
\label{Section: Conclusion}
An EVCI consisting of four CBs and six charging ports is simulated as shown in Fig. \ref{fig: EVCI as CPS}. The communications of the EVCI system are also simulated with suitable protocols like OCPP and GOOSE. The required data is obtained from the simulation and segregated into training and testing data as needed. A data-driven approach is used to predict the $\Delta SoC$, which adopts a ridge regression based ML model due to the multicollinearity between the data variables. This work assumes that anomalies are caused due to spoofing the communicated $\Delta SoC$ values. Absolute differences between the predicted $\Delta SoC$ and the communicated $\Delta SoC$ are computed and these differences are compared with the predefined threshold. The proposed methodology iteratively compares the absolute differences with the threshold to detect the anomalies. This work differentiates the anomalies with the EV's arrival and departure scenarios. 

Three types of spoofing techniques are implemented on the testing data and the proposed algorithm is validated. various case studies are evaluated to validate the proposed algorithm for the spoofing classification. The spoofing techniques are classified by observing the patterns in the calculated absolute differences. Spoofing $\Delta SoC$ values results in the irregular and improper charging of the EVs leading to energy theft, and financial loss to the EVCI administrators and the EV users. The possible future works are as follows:
\begin{itemize}
    \item Load forecasting of the EVCI as the manipulation of the SoC leads to performance degradation of batteries and causes irregular loading behavior on EVCI.
    \item Study of the grid instabilities due to large scale cyber attacks targeting the multiple EV's SoC impacting the power grid, leading to fluctuations and potential instabilities in the grid.
    \item Congestion management and handling the unplanned charging sessions that may arise due to false or manipulated SoC values.
    \item Estimating the revenue loss incurred by the energy theft due to the cyber attacks on the EVCI.
\end{itemize}

\bibliographystyle{IEEEtran}
\bibliography{main}

\end{document}